\documentclass{article}
\usepackage{amsmath}
\usepackage{bm}
\usepackage{graphicx}
\usepackage[dvipsnames]{xcolor}
\usepackage{hyperref}
\usepackage{relsize}
\usepackage[framemethod=tikz]{mdframed}
\usepackage{seqsplit}
\usepackage{float}
\usepackage{enumitem}
\usepackage{listings}
\definecolor{urlcolor}{rgb}{0.234, 0.234, 0.789} %url: blue
\definecolor{keycolor}{rgb}{0.5, 0, 0.5}  % keyword: dark magenta
\definecolor{comcolor}{rgb}{0, 0.5, 0} % comment: dark green
\title{Drawing ellipses and elliptical arcs with piecewise cubic Bézier curve approximations}
\author{Jerry R. Van\nobreak\hspace{.11em}Aken}
\begin{document}
  \maketitle

\begin{abstract}
\noindent This tutorial explains how to use piecewise cubic Bézier curves to draw arbitrarily oriented ellipses and elliptical arcs. The geometric principles discussed here result in strikingly simple interfaces for graphics functions that can draw (approximate) circles, ellipses, and arcs of circles and ellipses. C++ source code listings are included for these functions. Their code size can be relatively small because they are designed to be used with a graphics library or platform that draws Bézier curves, and the library or platform is tasked with the actual rendering of the curves. \\

\end{abstract}

%-------------------------------------------------------------------------------------------------------------------------------
\section{Introduction}

This tutorial explains how 2-D computer graphics functions of modest code size can use piecewise cubic B\'ezier curves to draw not only circles and circular arcs, but also ellipses and elliptical arcs of arbitrary shape and orientation.

Of course, a cubic polynomial curve segment cannot exactly represent an arc of a circle or ellipse, but the approximation error is relatively small and is easily adjusted to fit the needs of typical graphics applications.

Methods for approximating circles and circular arcs with B\'ezier curves have been extensively described in the literature [4][7][16][1] and on the Web [8][9][12]. Additionally, B\'ezier curves are \textit{affine invariant} [5][25][17]. Thus, if the control polygons for a piecewise B\'ezier curve are constructed to approximate a circle or circular arc, the vertices of these polygons can be affine-transformed to draw an ellipse or elliptical arc of arbitrary shape and orientation.

Typically, a 2-D graphics library specifies an ellipse in terms of the bounding rectangle in which the ellipse is inscribed [2][13][14]. This rectangle\footnote{The SVG \texttt{arc} command [24][23] and its implementations [10][11] require, as input parameters, the dimensions of the bounding rectangle but not the rectangle's center point, which is instead calculated to satisfy a set of constraints.} determines the principal axes of the ellipse. To draw a rotated ellipse, a library's ellipse function requires the caller to supply the angle of rotation as a function parameter [2][9][13][14]. The library's elliptical arc function specifies an arc of the ellipse with additional parameters, such as the arc's start and end points [10][11] or its start angle and sweep angle [2][9][13][14].

Inside these library functions, ellipses and elliptical arcs can be immediately converted to piecewise B\'ezier curves and added to the path. After that, the curves representing the arcs can be affine-transformed or otherwise manipulated in the same manner as the other curves in the path. Thus, for the library developer, at least, the consequences of dealing with whatever idiosyncratic and abstruse interfaces have been defined for the ellipse and elliptical arc functions are locally confined to the code for these functions and do not propagate into the library's path-rendering code to cause problems there.

The developer of a 2-D graphics application that calls the library's ellipse and elliptical arc functions may not be so fortunate. Any manipulation of ellipses and arcs performed by the application must take into account the requirements and limitations of the library's function interfaces. In particular, if the application needs to apply its own affine transformations to ellipses and arcs, special handling may be required to propagate these transformations into the parameters required by the library's ellipse and arc functions. For example:
\begin{itemize}
  \item What happens if an ellipse or arc of an ellipse must be defined in terms of an bounding rectangle, but the rectangle gets transformed into a parallelogram?
  \item If a direction flag indicates whether an arc is to be drawn in the clockwise or CCW direction, and the transformation incorporates a reflection that reverses the arc's direction from clockwise to CCW, or vice versa, how is this situation detected so that the flag can be flipped?
  \item If the library requires an ellipse's orientation to be specified as a rotation of the ellipse's bounding rectangle away from the $x$ or $y$ coordinate axis, how is the rotation angle to be retrieved from the application's internal representation of the transformed ellipse?
\end{itemize}
These problems are compounded if the application needs to be portable among several libraries or platforms.

In contrast, applying an affine transformation to a cubic B\'ezier curve is as simple as multiplying the four points that define the curve by the transformation matrix. The same graphics libraries that have problematic support for ellipses and arcs typically provide straightforward functions for drawing cubic B\'ezier curves. The interfaces to these B\'ezier functions consistently require, as input parameters, four points to define a curve segment.

What if a graphics application could bypass the library's ellipse and elliptical arc functions and convert its ellipses and arcs directly into B\'ezier curves?

This tutorial will present simple, intuitive interfaces for functions that use B\'ezier curves to draw ellipses and elliptical arcs. The C++ source code implementations of these functions are included in later sections. The relatively small code size of these implementations is due to their reliance on the graphics library's built-in B\'ezier curve functions to do most of the work. After all, the library's path-rendering apparatus typically handles stroking and filling the B\'ezier curves, subdividing and flattening the curves, clipping and antialiasing them, and perhaps even transforming them.

The simplicity of the function interfaces presented here follows from the fact that any ellipse can be uniquely described by just three points [22][21][20][19][18]. Applying an arbitrary affine transformation to these three points produces three new points that describe the transformed ellipse. No special handling is required. Consequently, an application that draws ellipses and elliptical arcs by calling the B\'ezier curve functions in one graphics library can typically be ported with relative ease to another library that supports B\'ezier curves.

For the sake of portability, graphics application developers might consider using the functions presented here as a work-around for whatever built-in support for circles, ellipses, and arcs is provided by the graphics libraries they are using.

%-------------------------------------------------------------------------------------------------------------------------------
\section{A simple interface for an ellipse function}

What is the simplest, most intuitive way for the user of a 2-D graphics application to specify an ellipse?

Any ellipse can be specified in terms of the square, rectangle, or parallelogram in which the ellipse is inscribed. As shown in \mbox{Figure 1}, the inscribed ellipse touches the midpoint of each of the four sides of the bounding square, rectangle, or parallelogram. At each midpoint, the side is tangent to the ellipse.

%////////////////////////////////////////////////////////////////
\begin{figure}[htb]
  \centering
  \fbox{%
    \begin{tabular}{p{0.9\textwidth}}
      \centering
      \includegraphics[width=10cm]{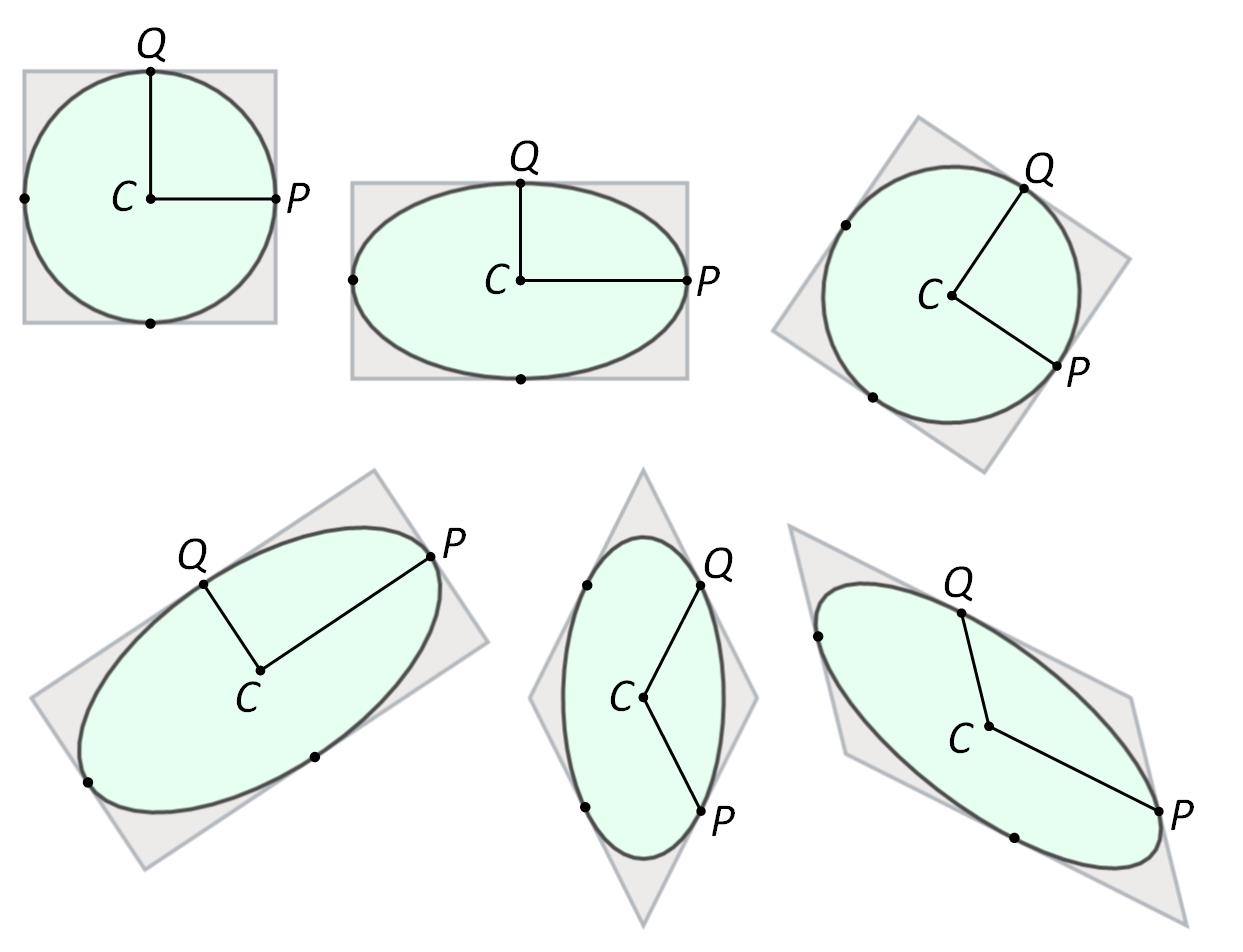} \\
      \caption{Ellipses inscribed in squares, rectangles, and parallelograms}
    \end{tabular}}
\end{figure}

%////////////////////////////////////////////////////////////////
\begin{figure}[htb]
  \centering
  \fbox{%
    \begin{tabular}{p{0.9\textwidth}}
      \centering
      \includegraphics[width=10cm]{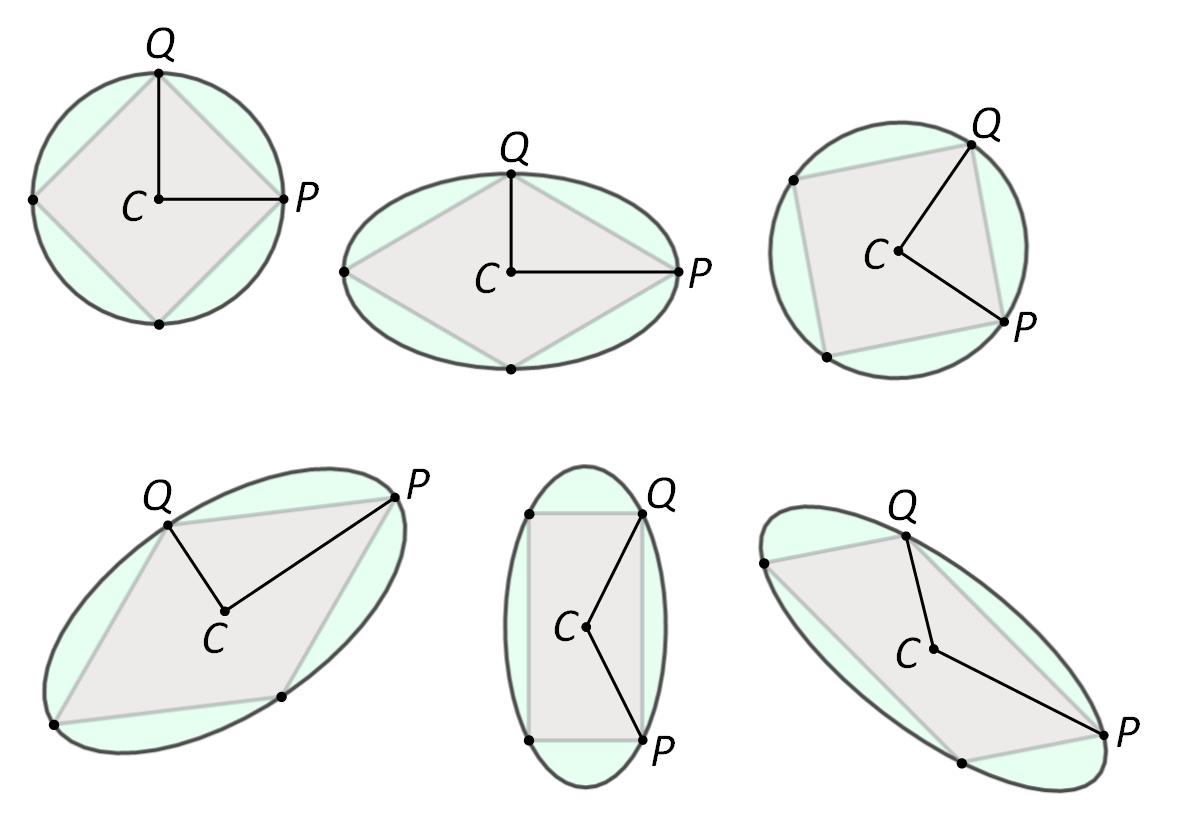} \\
      \caption{Squares, rectangles, and parallelograms inscribed in ellipses}
    \end{tabular}}
\end{figure}

Alternatively, any ellipse can be specified by the square, rectangle, or parallelogram that is inscribed \textit{inside} the ellipse. As shown in \mbox{Figure 2}, the circumscribing ellipse passes through each of the four vertices of the inscribed square, rectangle, or parallelogram.

In Figures 1 and 2, three points labeled $C$, $P$, and $Q$ are shown for each ellipse. These three points are sufficient to specify the position, shape, and orientation of any ellipse [22][21][20][19][18].

Point $C$ is the center of the ellipse.

In \mbox{Figure 1}, points $P$ and $Q$ are the midpoints of two adjacent sides of the square, rectangle, or parallelogram in which the ellipse is inscribed. $P$ and $Q$ are also end points of two \textit{conjugate diameters} of the ellipse\footnote{Two diameters of an ellipse are \textit{conjugate} if each diameter is parallel to the tangents at the ends of the other diameter.}. Observe that the tangent vector at point $P$ on an ellipse equals $Q-C$, and the tangent vector at point $Q$ equals $C-P$. The directions of these tangents are based on the assumption that the ellipse is drawn starting at $P$ and moving in the direction of $Q$.

If any three consecutive vertices $(V_0, V_1, V_2)$ of a bounding square, rectangle, or parallelogram in \mbox{Figure 1} are known, the fourth vertex $V_3$ is easily determined from symmetry. The inscribed ellipse can be described by the points
\begin{align*}
\\[-20pt]
  C = \frac12(V_0+V_2), \quad P = \frac12(V_0+V_1), \quad \textrm{and} \quad Q = \frac12(V_1+V_2)
\\[-20pt]
\end{align*}
Of course, conjugate diameter end points $P$ and $Q$ can be set to the midpoints of any pair of adjacent sides of the bounding parallelogram and still describe the same ellipse.

Similarly, in Figure 2, each circumscribing ellipse can be described by its center point, plus any pair of adjacent vertices of the inscribed parallelogram. These two vertices are end points of a pair of conjugate diameters of the ellipse.

These observations suggest using the following interface for a C++ function to draw ellipses:
\begin{lstlisting}[frame=none]
    void DrawEllipse(Point C, Point P, Point Q);
\end{lstlisting}
Parameter $C$ specifies the center of the ellipse, and $P$ and $Q$ are two conjugate diameter end points. Data type \texttt{Point} is a structure containing an $x$-$y$ coordinate pair that we can assume is defined by the graphics library that implements the B\'ezier curve function.

Aside comment: For the sake of accuracy, the input parameters to a graphics library's B\'ezier curve functions should allow the $x$-$y$ coordinates in the \texttt{Point} structure to be specified either as floating-point values or as fixed-point values. Forcing the vertices of B\'ezier control polygons to the nearest integer coordinates introduces up to a half pixel displacement error in either dimension. This error can significantly outweigh the error due to approximating the circle or ellipse with B\'ezier curves.

Ellipses drawn by the \texttt{DrawEllipse} function share an important property with Bézier curves: they are \textit{affine-invariant} [5][25][17]. For a Bézier curve, applying any affine transformation to the vertices $(p_1,c_1,c_2,p_2)$ of the curve's control polygon produces the same transformed curve as does individually transforming the points on the curve. Similarly, for an ellipse constructed by the \texttt{DrawEllipse} function, applying an affine transformation to the ellipse center point $C$ and the two conjugate diameter end points $P$ and $Q$ has the same effect as individually transforming the points on the ellipse.

Another useful property of the \texttt{DrawEllipse} function is that its input parameters contain all the information needed to determine a starting point (we'll use $P$) for drawing the ellipse and a direction (toward\footnote{That is, toward $Q$ along the shorter arc from $P$ to $Q$, and not the long way around.} $Q$) in which to start drawing. This property is important for graphics applications that require control over the direction in which points are plotted along a line, arc, or curve\,$-$\,for example, when constructing a shape that is to be filled according to the \textit{nonzero-winding-number} rule, as shown in \mbox{Figure 3}.

%////////////////////////////////////////////////////////////////
\begin{figure}[ht]
\centering
\includegraphics[width=10cm]{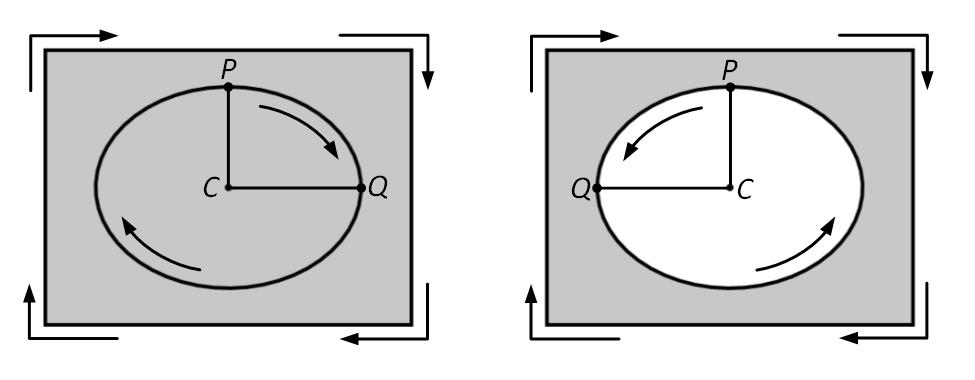}
\caption{Nonzero-winding-number fills}
\end{figure}

Consider a graphics library with a circle-drawing function that, for example, takes the circle radius and center coordinates as input parameters. The function has no way to determine from these parameters whether the caller intends to draw the circle in the clockwise or CCW direction. A library developer stuck with using such parameters can choose either to ignore this problem or to address it by providing an additional parameter, such as a direction flag.

In contrast, our \texttt{DrawEllipse} function provides the caller with straightforward direction control without additional parameters.

Additionally, if an affine transformation matrix with a reflection is applied to the \texttt{DrawEllipse} function's $C$, $P$, and $Q$ parameters, so that the ellipse is transformed to its mirror image, the resulting change in the relative positions of points $P$ and $Q$ automatically flips the drawing direction from clockwise to CCW, or vice versa. This property will prove to be especially important in the following discussion of the \texttt{DrawEllipticalArc} function.

The \texttt{DrawEllipse} function can be enhanced to draw arcs of circles and ellipses. The C++ declaration for the arc-drawing function is
\begin{lstlisting}[frame=none]
    void DrawEllipticalArc(Point C, Point P, Point Q,
                           float astart, float asweep);
\end{lstlisting}
Parameters $C$, $P$, and $Q$ describe the circle or ellipse containing the arc and have the same meaning as in the \texttt{DrawEllipse} function. The two new parameters, \texttt{astart} and \texttt{asweep}, specify the starting angle for the arc and the angle swept out by the arc. Both angles are expressed in radians. Either angle can be positive or negative.

For a circle, parameters \texttt{astart} and \texttt{asweep} are circular angles. For an ellipse, \texttt{astart} and \texttt{asweep} are affine mappings of circular angles onto the ellipse.

Starting angle \texttt{astart} is specified with respect to point $P$, with angles increasing toward $Q$ and decreasing in the opposite direction. Sweep angle \texttt{asweep} is positive or negative in the same directions as \texttt{astart}\footnote{If \texttt{astart} = 0, the arc begins at point $P$, and a positive \texttt{asweep} angle initiates drawing from $P$ in the direction of $Q$. If \texttt{asweep} = 0, no arc is drawn, but path-construction conventions might require the arc starting point to be added to the path (a detail left to the developer).}. These conventions ensure that affine transformations of elliptical arcs always behave as expected.

%////////////////////////////////////////////////////////////////
\begin{figure}[htb]
  \centering
  \fbox{%
    \begin{tabular}{p{0.9\textwidth}}
      \centering
      \includegraphics[width=11cm]{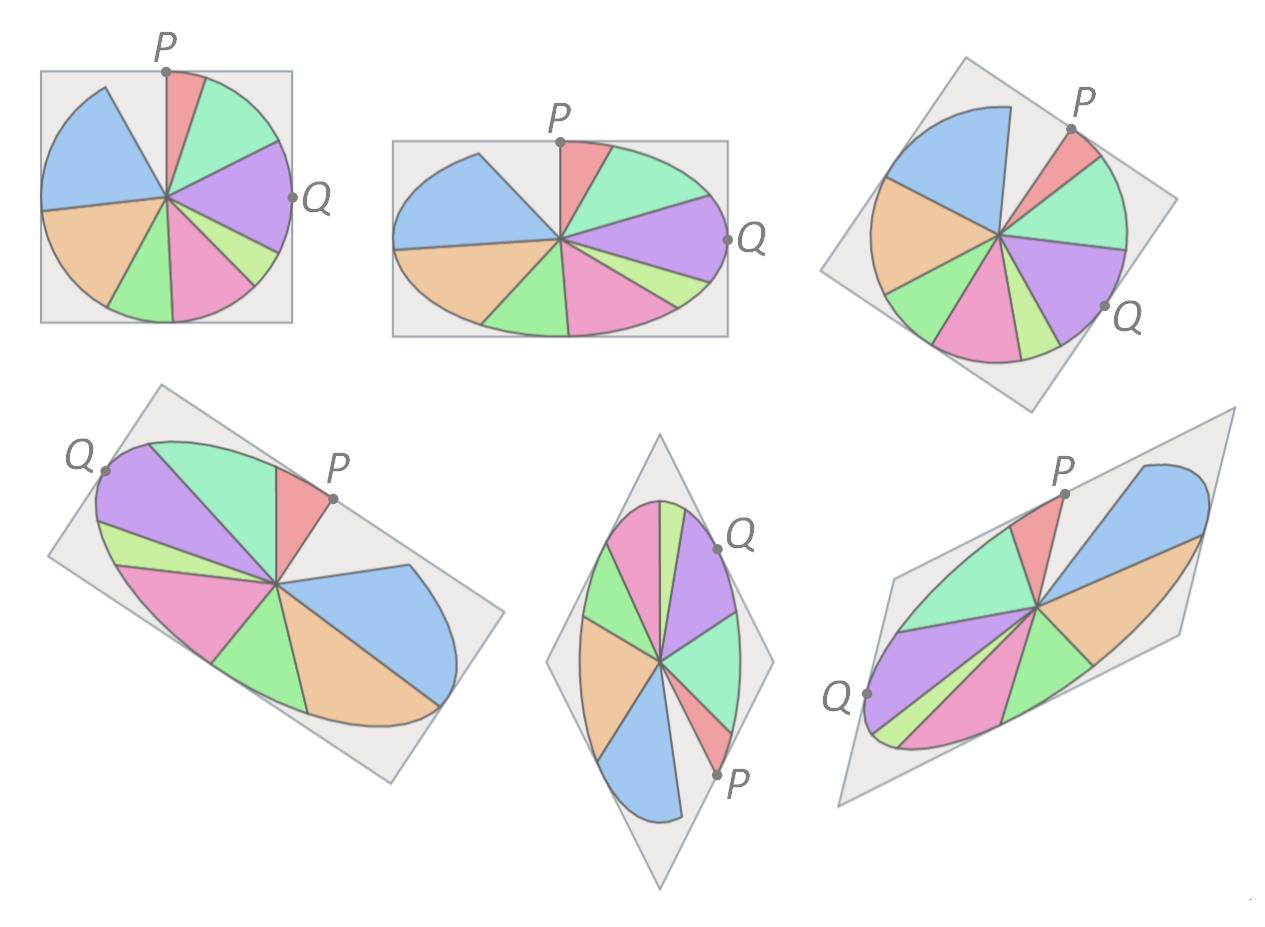} \\
      \caption{Pie charts drawn with elliptical arcs}
    \end{tabular}}
\end{figure}

Figure 4 shows six pie charts drawn with the \texttt{DrawEllipticalArc} function. All the arcs in a particular pie chart are part of the same ellipse, and so the $C$, $P$, and $Q$ parameters are identical for each of these arcs. Each arc's position on the ellipse is specified by a different set of \texttt{astart} and \texttt{asweep} parameters. The \texttt{astart} and \texttt{asweep} parameters for a particular pie slice (say, the orange one) are identical across all six pie charts. Thus, an affine transformation of a pie chart in the figure requires only that the points $C$, $P$, and $Q$ be transformed; no change is required to the \texttt{astart} and \texttt{asweep} values. A later section will explain how this effect is achieved.

In the top row of \mbox{Figure 4}, points $P$ and $Q$ for each pie chart are positioned so that the starting angle and sweep angle increase in the clockwise direction; in the bottom row, they increase in the CCW direction. This change in direction is achieved solely by changing the relative positions of points $P$ and $Q$. Again, the \texttt{astart} and \texttt{asweep} parameter values for a particular pie slice (such as the orange one) do not change from one pie chart to the next.

C++ source code listings for the \texttt{DrawEllipse} and \texttt{DrawEllipticalArc} functions will be presented in a later section. But first we will discuss (1) how to construct a cubic B\'ezier curve segment that closely approximates an arc of a circle, and (2) how to map the resulting B\'ezier control points from the circle to an ellipse.

%-------------------------------------------------------------------------------------------------------------------------------
\section{Approximating a circular arc}

This section explains how to construct a B\'ezier curve segment that approximates a circular arc.

To draw a full circle, partition it into multiple arcs of uniform size; then use a B\'ezier curve segment to approximate each arc. To partition the circle into four arcs, for example, generate arc end points around the circle at intervals of $\pi/2$ radians.

Figure 5 shows a cubic B\'ezier curve segment being used to approximate\footnote{In Figure 5, the B\'ezier curve diverges slightly from the circular arc at points between the arc midpoint and the two ends of the arc, but this divergence is too small to be depicted in the figure.} an arc on a circle of radius $R$. The circle is centered at coordinate origin $O$. The arc spans central angle $\phi$.

%////////////////////////////////////////////////////////////////
\begin{figure}[htb]
  \centering
  \fbox{%
    \begin{tabular}{p{0.9\textwidth}}
      \centering
      \includegraphics[width=8cm]{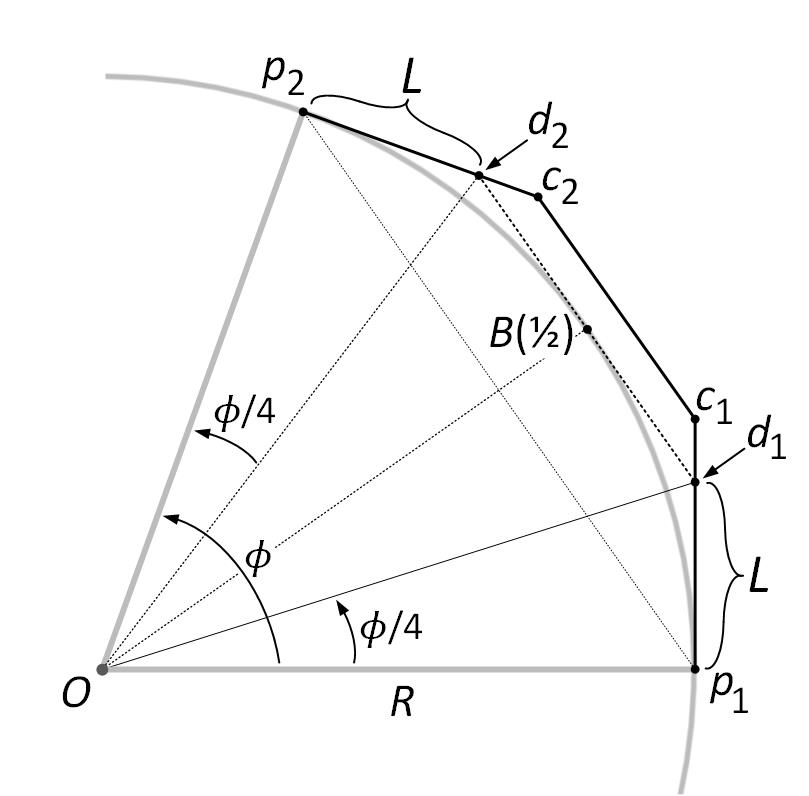} \\
      \caption{A circular arc approximated by the cubic B\'ezier curve segment from $p_1$  to $p_2$}
    \end{tabular}}
\end{figure}

The B\'ezier curve in the figure is defined by the four points $(p_1,c_1,c_2,p_2)$. The curve begins at $p_1$ and ends at $p_2$. Between these two points, the curve is pulled toward (but does not interpolate) control points $c_1$ and $c_2$.

The B\'ezier control polygon in \mbox{Figure 5} is bilaterally symmetrical across the radial line from the origin to the midpoint of the arc. Line segments $p_1.p_2$ and $c_1.c_2$ are parallel to the tangent at the arc midpoint.

This tutorial will use a well-known strategy [4][7][8] for approximating a circular arc, which is to place B\'ezier points $p_1$ and $p_2$ at the ends of the arc, and to position $c_1$ and $c_2$ so that the B\'ezier curve interpolates the midpoint of the arc. Additionally, $c_1$ and $c_2$ are placed on the tangents to the circle at $p_1$ and $p_2$, respectively. The problem then is to determine exactly where on these two tangents to place $c_1$ and $c_2$ so that the curve passes through the arc midpoint.

The equation for a cubic B\'ezier curve is
\begin{align*}
B(t) = (1-t)^3 p_1 + 3t(1-t)^2 c_1 + 3t^2(1-t) c_2 + t^3 p_2
\end{align*}
where independent parameter $t$ ranges from $0$ at $p_1$ to $1$ at $p_2$. The point located at the middle of this curve, where $t=1/2$, is
\begin{equation}
B(\frac12) = \frac18 (p_1 + 3c_1 + 3c_2 + p_2)
\end{equation}

In \mbox{Figure 5}, control points $c_1$ and $c_2$ have been positioned so that the point $B(\frac12)$ coincides with the midpoint of the circular arc. The tangent to the circle at $B(\frac12)$ is shown intersecting with the tangents at $p_1$ and $p_2$. The points of intersection are labeled $d_1$ and $d_2$.

By inspecting \mbox{Figure 5}, it may be apparent to readers familiar with \textit{de Casteljau subdivision} [6][3] that $d_1$ is located 3/4 of the way from $p_1$ to $c_1$, and that $d_2$ is located 3/4 of the way from $p_2$ to $c_2$.

However, these facts can also be verified by observing that the point $B(\frac12)$ is located midway between $d_1$ and $d_2$, and that the equation $B(\frac12)=\frac12(d_1+d_2)$, where $d_1=\frac14 p_1+\frac34 c_1$ and $d_2=\frac14 p_2+\frac34 c_2$, is identical to \mbox{equation (1)}.

In \mbox{Figure 5}, the lengths $|p_1.d_1|$ and $|p_2.d_2|$ are labeled $L$. The lengths $|p_1.c_1|$ and $|p_2.c_2|$ must therefore be $\frac43L$.

Observe that the radial line from the circle center to $p_1$ forms a right angle with the tangent line $p_1.d_1$, and that these lines form the two legs of a right triangle. The angle that is opposite side $p_1.d_1$ of this triangle is $\phi/4$, so that $L/R=\tan\frac\phi4$, and $L = R\tan\frac\phi4$.

We now know that the lengths $|p_1.c_1|$ and $|p_2.c_2|$ equal $\frac43L=R(\frac43\tan\frac\phi4)$. The vectors $c_1-p_1$ and $c_2-p_2$ can be obtained by multiplying the tangent vectors at points $p_1$ and $p_2$ by $\frac43\tan\frac\phi4$, as described in [4][7][8]. For a circle, these tangents are easily calculated as perpendiculars to the radial lines from the circle center to $p_1$ and $p_2$. This approach fails, of course, in the general case in which the circle has been affine-transformed into an ellipse.

For the ellipse, the transformed radial lines and their associated tangents might no longer be perpendicular and of equal length. However, we can still obtain these tangents by exploiting the fact that parallel lines remain parallel under affine transformations, and that the ratio of the lengths of two parallel line segments are also invariant [17].

\setlength{\skip\footins}{0.4cm}
Recall from an earlier discussion that if $P$ and $Q$ are end points of a pair of conjugate diameters of an ellipse centered at point $C$, the tangent vector\footnote{As previously discussed, the directions of tangents to the ellipse are based on the assumption that the ellipse is drawn starting at $P$ and moving toward $Q$.} at point $P$ on the ellipse equals $Q-C$. For simplicity, assume that this ellipse is centered at the origin so that $Q-C=Q$. If $P$ is placed at starting point $p_1$ of the transformed B\'ezier curve on the ellipse, then
\\[-3pt]
\begin{equation}
c_1-p_1 = Q(\frac43\tan\frac{\phi}4) \qquad\;\;\; \text{where} \;\: p_1 = P
\end{equation}

As will be discussed shortly, points $P$ and $Q$ can be rotated in unison around the ellipse to positions $P'$ and $Q'$ that are end points of a new pair of conjugate diameters on the same ellipse. If this rotation places $P'$ at point $p_2$ on the B\'ezier curve, the tangent vector at $p_2$ is equal to $Q'$, but is pointing \textit{away} from $c_2$. Taking into account the resulting sign change, we have
\begin{equation}
c_2-p_2 = -Q'(\frac43\tan\frac{\phi}4) \qquad\;\;\; \text{where} \;\: p_2 = P'
\end{equation}

In summary, equations (2) and (3) can be used to calculate B\'ezier control points $c_1$ and $c_2$ for an (approximate) arc of an ellipse given the appropriately positioned end points $P$ and $Q$ of a pair of conjugate diameters of the ellipse. The ellipse is of arbitrary shape and orientation, and is centered at the origin.

%-------------------------------------------------------------------------------------------------------------------------------
\section{Transforming points on the unit circle to an ellipse}

The previous section explained how to position the four points $(p_1,c_1,c_2,p_2)$ that define a cubic B\'ezier curve so that the curve closely approximates a circular arc. This section will discuss how to transform points on a circle to points on an ellipse, and how rotations around the circle map to rotations around the ellipse [22][21][20][19].

To partition a circle into four arcs, for example, arc end points are generated around the circle at intervals of $\pi/2$ radians. This process is pretty straightforward. But if we intend to map the arcs on the circle to arcs on an arbitrarily oriented ellipse, we must first determine the affine transformation from the points on the circle to the ellipse.

\mbox{Figure 6} shows that an ellipse and its bounding parallelogram are the affine-transformed images of a unit circle and its bounding square. We use $u$-$v$ coordinates for the unit circle, and $x$-$y$ coordinates for the ellipse. The circle and ellipse are centered at their respective origins.

%////////////////////////////////////////////////////////////////
\begin{figure}[htb]
  \centering
  \fbox{%
    \begin{tabular}{p{0.9\textwidth}}
      \centering
      \includegraphics[width=9.5cm]{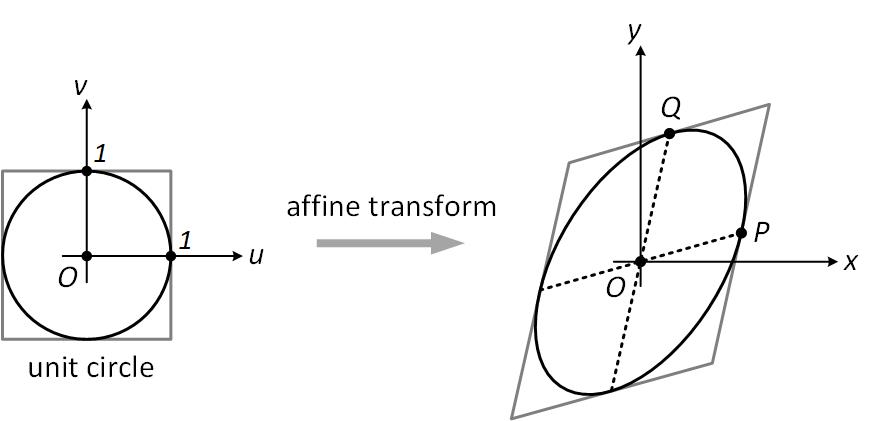} \\
      \caption{Affine transformation of a unit circle and its bounding square to an ellipse and its bounding parallelogram}
    \end{tabular}}
\end{figure}

In \mbox{Figure 6}, conjugate diameter end points $(1,0)$ and $(0,1)$ on the unit circle are transformed to points $P = (x_P,y_P)$ and $Q = (x_Q,y_Q)$ on the ellipse. We know that
\begin{itemize}
  \item Any pair of perpendicular diameters of a circle are conjugate diameters of the circle.
  \item An affine transformation of a circle to an ellipse will map any pair of conjugate diameters of the circle to the corresponding pair of conjugate diameters of the ellipse.
\end{itemize}
Thus, points $P$ and $Q$ in \mbox{Figure 6} lie on conjugate diameters of the ellipse.

The affine transformation of a point $(u,v)$ on the unit circle to a point $(x,y)$ on the origin-centered ellipse can be expressed as
\begin{align*}
\begin{bmatrix} x \\  y \end{bmatrix}
= \mathbf{M}
\begin{bmatrix} u \\ v \end{bmatrix}
\qquad\:\textrm{where}\:\:\mathbf{M} =
\begin{bmatrix} m_{11}  & m_{12}  \\ m_{21}  & m_{22} \end{bmatrix}
\end{align*}
To determine the coefficients $m_{ij}$ of transformation matrix $\mathbf{M}$, consider that conjugate diameter end points $P$ and $Q$ on the ellipse are related to points $(1,0)$ and $(0,1)$ on the unit circle by the expressions $P = \mathbf{M} \mathsmaller{\begin{bmatrix} 1 & 0 \end{bmatrix}^{\mathsf{T}} }$ and $Q = \mathbf{M} \mathsmaller{\begin{bmatrix} 0 & 1 \end{bmatrix}^{\mathsf{T}} }$. These two expressions expand to the following:
\begin{align*}
\\[-14pt]
\begin{bmatrix}
x_P \\ y_P
\end{bmatrix}
& =
\begin{bmatrix}
m_{11} & m_{12} \\
m_{21} & m_{22}
\end{bmatrix}
\begin{bmatrix}
1 \\ 0
\end{bmatrix}   \\[+2pt]
& =
\begin{bmatrix}
m_{11} \\ m_{21}
\end{bmatrix}
\\[+8pt]
\begin{bmatrix}
x_Q \\ y_Q
\end{bmatrix}
& =
\begin{bmatrix}
m_{11} & m_{12} \\
m_{21} & m_{22}
\end{bmatrix}
\begin{bmatrix}
0 \\ 1
\end{bmatrix}   \\[+2pt]
& =
\begin{bmatrix}
m_{12} \\ m_{22}
\end{bmatrix}
\end{align*}
From inspection, we see that
\begin{equation*}
\mathbf{M} =
\begin{bmatrix}
x_P & x_Q \\
y_P & y_Q
\end{bmatrix}
\end{equation*}
The affine transformation of a point $(u,v)$ on the unit circle to the corresponding point $(x,y)$ on the ellipse can now be expressed as
\begin{align}
\begin{bmatrix}
x \\ y
\end{bmatrix}
& =
\begin{bmatrix}
x_P & x_Q \\
y_P & y_Q
\end{bmatrix}
\begin{bmatrix}
u \\ v
\end{bmatrix}
\end{align}
where the coefficients of the 2$\times$2 matrix are the $x$ and $y$ coordinates of end points $P$ and $Q$ of a pair of conjugate diameters of the ellipse [22][21].

If, on the left side of \mbox{Figure 6}, the unit circle's conjugate parameter end points $(1,0)$ and $(0,1)$ are rotated in unison by the same angle $\phi$, the coordinates of the rotated points are $(\cos\phi,\sin\phi)$ and $(-\sin\phi,\cos\phi)$. The diameters through these points are perpendicular and thus are new conjugate diameters of the circle.

If \mbox{equation (4)} is then used to transform the rotated points on the unit circle to the corresponding points on the ellipse, the transformed points $P'=(x_P',y_P')$ and $Q'=(x_Q',y_Q')$ will lie on a new pair of conjugate diameters of the same ellipse [22][21]. The transformations of these two points can be expressed as
\begin{align*}
\begin{bmatrix} x_P' \\ y_P' \end{bmatrix}
=
\begin{bmatrix} x_P & x_Q \\ y_P & y_Q \end{bmatrix}
\begin{bmatrix} \cos\phi \\ \sin\phi \end{bmatrix}
\qquad \text{and} \qquad
\begin{bmatrix} x_Q' \\ y_Q' \end{bmatrix}
=
\begin{bmatrix} x_P & x_Q \\ y_P & y_Q \end{bmatrix}
\begin{bmatrix} -\sin\phi \\ \cos\phi \end{bmatrix}
\end{align*}
For brevity's sake, these two equations can be consolidated as follows:
\begin{align}
\\[-14pt]
\begin{bmatrix}
x_P' & x_Q' \\ y_P' & y_Q'
\end{bmatrix}
& =
\begin{bmatrix}
x_P & x_Q \\
y_P & y_Q
\end{bmatrix}
\begin{bmatrix}
\cos \phi & -\sin \phi \\
\sin \phi & \cos \phi
\end{bmatrix} \notag
\end{align}
This matrix equation can be used to rotate any pair of conjugate diameter end points $P$ and $Q$ on an origin-centered ellipse to a new pair of conjugate diameter end points $P'$ and $Q'$ on the same ellipse.

The rightmost matrix in \mbox{equation (5)} is easily recognizable as a proper rotation matrix. As previously discussed, this matrix could be the result a single rotation by angle $\phi$ away from the initial coordinates $(1,0)$ and $(0,1)$ on the unit circle. More generally, however, the matrix also could represent the accumulation of a number of rotations by various angles that sum to $\phi$.

After conjugate diameter end points $P$ and $Q$ have been rotated by angle $\phi$ to new end points $P'$ and $Q'$, as in \mbox{equation (5)}, rotating $P'$ and $Q'$ by a second angle $\alpha$ has the same effect as rotating $P$ and $Q$ by the sum ($\phi + \alpha$) of the two angles.

In equation (5), the rotation matrix appears to the right of the matrix containing the coordinates of points $P$ and $Q$. If the ordering of these two matrices were reversed, the transformation would simply rotate $P$ and $Q$ around a pair of circles centered at the origin. That's the way that rotation matrices are typically used but not what we want in this case.

Instead, we want to map the rotation of a pair of conjugate diameter end points around the unit circle to the corresponding rotation of a pair of conjugate diameter end points around an ellipse. \mbox{Equation (5)} provides this mapping.

%-------------------------------------------------------------------------------------------------------------------------------
\section{The \texttt{DrawEllipse} function}

The following C++ source code listing is an implementation of the \texttt{DrawEllipse} function discussed in \mbox{Section 2}. This function draws the circle or ellipse that is defined by parameters $C$, $P$, and $Q$, where $C$ is the center point of the ellipse, and $P$ and $Q$ are end points of a pair of conjugate diameters of the ellipse.

\begin{lstlisting}[caption=Draw ellipse composed of five arcs of uniform size $\frac25\pi$ radians]
  void DrawEllipse(Point C, Point P, Point Q)
  {
      const int nsegs = 5;
      const float PI = 3.141592653589793;
      const float phi = 2*PI/nsegs;
      const float cosp = cos(phi), sinp = sin(phi);
      const float tau = (4.0/3)*tan(phi/4);
      Point tmp, p1, c1, c2, p2;

      // Save the starting point for the ellipse
      p2 = P;

      // Convert conjugate diameter end points P and Q to
      // center-relative coordinates
      P.x -= C.x;
      P.y -= C.y;
      Q.x -= C.x;
      Q.y -= C.y;

      // Calculate the initial Bezier control point
      c2.x = p2.x - tau*Q.x;
      c2.y = p2.y - tau*Q.y;

      // For each elliptical arc of 'phi' radians, plot the
      // Bezier curve segment that approximates the arc
      for (int i = 0; i < nsegs; ++i)
      {
          p1 = p2;
          c1.x = p1.x + (p1.x - c2.x);
          c1.y = p1.y + (p1.y - c2.y);
          tmp.x = P.x*cosp + Q.x*sinp;
          tmp.y = P.y*cosp + Q.y*sinp;
          Q.x = Q.x*cosp - P.x*sinp;
          Q.y = Q.y*cosp - P.y*sinp;
          P = tmp;
          p2.x = P.x + C.x;
          p2.y = P.y + C.y;
          c2.x = p2.x - tau*Q.x;
          c2.y = p2.y - tau*Q.y;
          CubicBezier(p1, c1, c2, p2);
      }
  }
\end{lstlisting}

The \texttt{DrawEllipse} function is designed to be used with a graphics library that supports B\'ezier curves. The function uses a data type \texttt{Point} that we can assume is defined by the library. This type denotes a data structure containing a pair of $x$-$y$ coordinates. The coordinates should be either floating-point or fixed-point values. Integer coordinates can cause significant loss of precision in the calculations performed by the function.

The graphics library implements the \texttt{CubicBezier} function that is called at the bottom of the \texttt{DrawEllipse} listing. \texttt{CubicBezier} draws the cubic B\'ezier curve defined by the four points $(p_1,c_1,c_2,p_2)$.

The \texttt{sin}, \texttt{cos}, and \texttt{tan} functions called by the \texttt{DrawEllipse} function are declared in the \texttt{math.h} header file\footnote{The \texttt{ceil} function called in the \texttt{DrawEllipticalArc} listing in the next section is also declared in \texttt{math.h}.} in the Standard C Library [15].

The \texttt{DrawEllipse} function partitions the ellipse into five uniformly sized elliptical arcs, which are approximated as cubic B\'ezier curve segments. Each iteration of the \texttt{for}-loop at the end of the listing draws one $\frac25\pi$-radian arc of the ellipse.

Inside this \texttt{for}-loop, \mbox{equation (5)} from the preceding section is used to rotate conjugate diameter end points $P$ and $Q$ by $\phi=\frac25\pi$ radians to the start of the next segment. This equation might be easier to recognize in the following form:
\begin{align*}
\\[-14pt]
\begin{bmatrix}
x_P' & x_Q' \\ y_P' & y_Q'
\end{bmatrix}
& =
\begin{bmatrix}
x_P \cos\phi + x_Q \sin\phi & x_Q \cos\phi - x_P \sin\phi \\
y_P \cos\phi + y_Q \sin\phi & y_Q \cos\phi - y_P \sin\phi
\end{bmatrix}
\end{align*}

Equation (3) from an earlier section is used to calculate control point $c_2$ in the \texttt{DrawEllipse} function. Near the top of the listing, constant \texttt{tau} is set to the quantity $\frac43\tan\frac\phi4$ that appears in this equation.

Note that each pair of adjacent B\'ezier curve segments in the circle or ellipse meet at a common point that is both the end point $p_2$ of the first segment and the starting point $p_1$ of the second segment. After $c_2$ for the first segment is calculated, $c_1$ for the second segment can be obtained simply by reflecting $c_2$ through the common point\footnote{Equation (2) could have been used to calculate $c_1$, but using reflection instead eliminates two multiplication operations from the \texttt{for}-loop body.}, which is located midway between the two control points (all three points lie on the same tangent line).

The constant \texttt{nsegs}, which is defined near the top of the \texttt{DrawEllipse} function, specifies the number of B\'ezier curve segments used to construct the ellipse. The interested reader can adjust this constant to experiment with the accuracy of the B\'ezier curve approximation to the circle or ellipse as the number of segments is varied.

A simpler version of the \texttt{DrawEllipse} function can be created by keeping the value of \texttt{nsegs} fixed. For the special case \texttt{nsegs} = 4, with the resulting arc size $\phi=\pi/2$, the sine and cosine terms can be replaced with constants $\sin\phi=1$ and $\cos\phi=0$. Also, the value of \texttt{tau} is fixed and can be set to numerical constant $\frac43\tan\frac\phi4 = 0.5522847498307934$. (This constant is, BTW, a ``magic number" that appears in various discussions of B\'ezier circles on the Web\footnote{Sometimes it appears in the form $\kappa = \frac43(\sqrt2-1)$.}.) The following C++ source code listing for the \texttt{DrawEllipse2} function is a revised version of the \texttt{DrawEllipse} function that incorporates these simplifications.
\\  % <--- May want to delete this if text is revised
\begin{lstlisting}[caption=Draw ellipse composed of four arcs of uniform size $\pi/2$ radians]
  void DrawEllipse2(Point C, Point P, Point Q)
  {
      const float tau = 0.5522847498307934;
      Point tmp, p1, c1, c2, p2;

      p2 = P;
      P.x -= C.x;
      P.y -= C.y;
      Q.x -= C.x;
      Q.y -= C.y;
      c2.x = p2.x - tau*Q.x;
      c2.y = p2.y - tau*Q.y;
      for (int i = 0; i < 4; ++i)
      {
          p1 = p2;
          c1.x = p1.x + (p1.x - c2.x);
          c1.y = p1.y + (p1.y - c2.y);
          tmp = Q;
          Q.x = -P.x;
          Q.y = -P.y;
          P = tmp;
          p2.x = P.x + C.x;
          p2.y = P.y + C.y;
          c2.x = p2.x - tau*Q.x;
          c2.y = p2.y - tau*Q.y;
          CubicBezier(p1, c1, c2, p2);
      }
  }
\end{lstlisting}

Another special case that produces a simplified version of \texttt{DrawEllipse} is \texttt{nsegs} = 8. For the resulting arc size, $\phi=\frac14\pi$, the sine and cosine terms are equal: $\sin\phi = \cos\phi = \frac12\sqrt2$. The following C++ source code listing for the \texttt{DrawEllipse3} function exploits this fact to eliminate four multiplications from the \texttt{for}-loop body. Also, the value of \texttt{tau} is fixed and can be set to numerical constant $\frac43\tan\frac\phi4 = 0.2652164898395440$.

\begin{lstlisting}[caption=Draw ellipse composed of eight arcs of uniform size $\pi/4$ radians]
  void DrawEllipse3(Point C, Point P, Point Q)
  {
      const float tau = 0.2652164898395440;
      const float sincos = 0.7071067811865475;
      Point tmp, p1, c1, c2, p2;
      p2 = P;
      P.x -= C.x;
      P.y -= C.y;
      Q.x -= C.x;
      Q.y -= C.y;
      c2.x = p2.x - tau*Q.x;
      c2.y = p2.y - tau*Q.y;
      for (int i = 0; i < 8; ++i)
      {
          p1 = p2;
          c1.x = p1.x + (p1.x - c2.x);
          c1.y = p1.y + (p1.y - c2.y);
          tmp.x = sincos*(P.x + Q.x);
          tmp.y = sincos*(P.y + Q.y);
          Q.x = sincos*(Q.x - P.x);
          Q.y = sincos*(Q.y - P.y);
          P = tmp;
          p2.x = P.x + C.x;
          p2.y = P.y + C.y;
          c2.x = p2.x - tau*Q.x;
          c2.y = p2.y - tau*Q.y;
          CubicBezier(p1, c1, c2, p2);
      }
  }
\end{lstlisting}

Note that all three versions of \texttt{DrawEllipse} avoid imposing arbitrary restrictions on the input parameters supplied by the caller. Any three points $C$, $P$, and $Q$ uniquely define a circle or ellipse.

%-------------------------------------------------------------------------------------------------------------------------------
\section{The \texttt{DrawEllipticalArc} function}

The second function discussed in \mbox{Section 2} is the \texttt{DrawEllipticalArc} function, which draws an arc of a circle or ellipse. The following C++ source code listing is an implementation of this function. In the function header, parameters $C$, $P$, and $Q$ specify a circle or ellipse, and have the same meaning as in the \texttt{DrawEllipse} function. Parameters \texttt{astart} and \texttt{asweep} are, respectively, the starting angle for the arc and the angle swept out by the arc.

\begin{lstlisting}[caption=Draw elliptical arc with maximum subarc size of $\pi/2$ radians]
  void DrawEllipticalArc(Point C, Point P, Point Q,
                         float astart, float asweep)
  {
      const float PI = 3.141592653589793;
      const float maxphi = PI/2;
      Point tmp, p1, c1, c2, p2;

      if (asweep == 0)
          return;  // zero-length arc

      // Convert conjugate diameter end points P and Q to
      // center-relative coordinates
      P.x -= C.x;
      P.y -= C.y;
      Q.x -= C.x;
      Q.y -= C.y;

      // Generate new conjugate diameter end points P' and Q'
      // by rotating points P and Q by starting angle astart
      if (astart != 0)
      {
          float cosa = cos(astart), sina = sin(astart);
          tmp.x = P.x*cosa + Q.x*sina;
          tmp.y = P.y*cosa + Q.y*sina;
          Q.x = Q.x*cosa - P.x*sina;
          Q.y = Q.y*cosa - P.y*sina;
          P = tmp;
      }

      // Set the starting point for the elliptical arc
      p2.x = P.x + C.x;
      p2.y = P.y + C.y;

      // If the sweep angle is negative, convert it to a
      // positive angle pointing in the opposite direction
      if (asweep < 0)
      {
          Q.x = -Q.x;
          Q.y = -Q.y;
          asweep = -asweep;
      }
      if (asweep > 2*PI)
          asweep = 2*PI;

      // If the arc's sweep angle is too big to be accurately
      // drawn as a single Bezier curve segment, partition it
      // into smaller angles of uniform size 'phi'
      int nsegs = 1;
      float phi = asweep;
      if (asweep > maxphi)
      {
          nsegs = ceil(asweep/maxphi);
          phi = asweep/nsegs;
      }

      // Use tuning parameter 'tau' to calculate initial
      // Bezier control point c2
      float tau = (4.0f/3)*tan(phi/4);
      c2.x = p2.x - tau*Q.x;
      c2.y = p2.y - tau*Q.y;

      // For each elliptical arc of 'phi' radians, plot a
      // Bezier curve segment to approximate the arc
      float cosp = cos(phi), sinp = sin(phi);
      for (int i = 0; i < nsegs; ++i)
      {
          p1 = p2;
          c1.x = p1.x + (p1.x - c2.x);
          c1.y = p1.y + (p1.y - c2.y);
          tmp.x = P.x*cosp + Q.x*sinp;
          tmp.y = P.y*cosp + Q.y*sinp;
          Q.x = Q.x*cosp - P.x*sinp;
          Q.y = Q.y*cosp - P.y*sinp;
          P = tmp;
          p2.x = P.x + C.x;
          p2.y = P.y + C.y;
          c2.x = p2.x - tau*Q.x;
          c2.y = p2.y - tau*Q.y;
          CubicBezier(p1, c1, c2, p2);
      }
  }
\end{lstlisting}

The \texttt{DrawEllipticalArc} function body is similar that of the \texttt{DrawEllipse} function\,$-$\,in fact, their \texttt{for}-loops are identical (refer to the first \texttt{DrawEllipse} listing, not the \texttt{DrawEllipse2} or \texttt{DrawEllipse3} version).

However, \texttt{DrawEllipticalArc} must handle two additional input parameters, \texttt{astart} and \texttt{asweep}. If \texttt{astart} is nonzero, conjugate diameter end points $P$ and $Q$ must be rotated to their starting positions. Also, the function must determine whether the arc specified by sweep angle \texttt{asweep} is small enough to be approximated with sufficient accuracy by a single B\'ezier curve segment.

An \texttt{asweep} angle that is too big to be spanned by a single B\'ezier curve segment is partitioned into two or more smaller angles of uniform size \texttt{phi}. The \texttt{maxphi} constant at the top of the function body defines the maximum size angle that can be spanned by a curve segment. The interested reader can adjust this constant to experiment with the accuracy of the piecewise B\'ezier curve approximation to the elliptical arc for different maximum angle values.

Note that \texttt{DrawEllipticalArc} avoids imposing arbitrary restrictions on the input parameters supplied by the caller. Any three points $C$, $P$, and $Q$ uniquely define a circle or ellipse, and the \texttt{astart} and \texttt{asweep} parameters can each be set to any angle in the range $-2\pi$ to $+2\pi$ and produce the expected result.

%-------------------------------------------------------------------------------------------------------------------------------
\section{Approximation error}

As previously mentioned, a cubic B\'ezier curve segment cannot exactly represent a circular arc. Between the arc midpoint and either arc end point in \mbox{Figure 5}, the curve drifts outside the circle that contains the arc. For an arc with a reasonably sized central angle $\phi$, the resulting error is relatively small and might be difficult to discern on a typical graphics display.

To keep the approximation error small, graphics applications typically keep the arc's central angle from exceeding a maximum size set somewhere in the range $\pi/4$ to $\pi/2$ radians\footnote{If the \texttt{DrawEllipticalArc} function in \mbox{Section 4} is asked to draw an arc that exceeds the size limit, the function partitions the arc into two or more subarcs that meet the size limit.}. For a circle with a radius of a thousand pixels that is segmented into four arcs of $\pi/2$ radians, the maximum radial error is about a quarter of a pixel. If the same circle is segmented into eight $\pi/4$-radian arcs, the maximum error shrinks to a tiny 1/236 of a pixel.

%////////////////////////////////////////////////////////////////
\begin{figure}[htb]
  \centering
  \fbox{%
    \begin{tabular}{p{0.9\textwidth}}
      \centering
      \includegraphics[width=9.5cm]{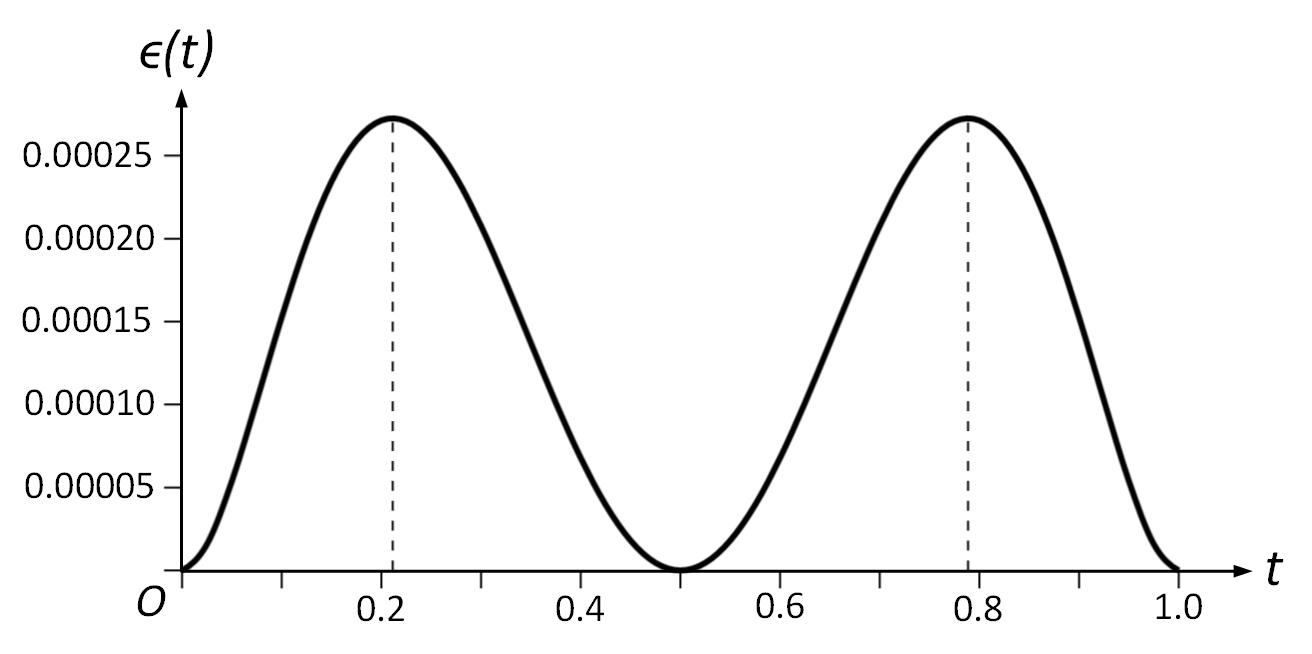} \\
      \caption{Error $\epsilon(t)$ (Euclidean distance) over a B\'ezier curve segment $B(t)$, $t \in [0,1]$, that approximates an arc of $\phi=\pi/2$ radians on the unit circle}
    \end{tabular}}
\end{figure}

Figure 7 shows the radial error in the B\'ezier curve approximation of an arc of $\pi/2$ radians on the unit circle. The B\'ezier curve approximation used for this analysis was previously described in \mbox{Section 3}, and uses \mbox{equations (2)} and (3) to calculate B\'ezier control points $c_1$ and $c_2$. The error $\epsilon(t)$ is calculated as the Euclidean distance from the B\'ezier curve to the unit circle. That is, given the coordinates $x(t)$ and $y(t)$ of a point on the B\'ezier curve, the error is
\begin{align*}
\epsilon(t) = \sqrt{x^2(t)+y^2(t)}-1
\end{align*}
Because the B\'ezier curve never drifts \textit{inside} the circle containing the arc, we have that $\sqrt{x^2(t)+y^2(t)} \ge 1$ and $\epsilon(t) \ge 0$ for $t\in[0,1]$.

The two dashed lines in \mbox{Figure 7} indicate where the maximum errors occur. For our B\'ezier curve approximation of an arc of length $\phi$ on the unit circle, Dokken \textit{et al} [4] determine that the \textit{non}-Euclidean error function
\begin{align*}
\psi(t) = | x^2(t) + x^2(t) - 1 |
\end{align*}
has the maximum value
\\[-18pt]
\begin{align*}
\psi_{max}(\phi) = \max_{t\in[0,1]} \{\psi(t)\} = \frac{4}{27}\frac{\sin^6 \frac\phi4}{\cos^2\frac\phi4}
\end{align*}
at $t = (3\pm\sqrt3)/6$. The error functions $\psi(t)$ and $\epsilon(t)$ are related as follows [4]:
\begin{align*}
\psi(t) &= \big( \sqrt{x^2(t)+y^2(t)}-1 \big) \big(\sqrt{x^2(t)+y^2(t)}+1 \big) \\
         &= \epsilon(t) \big(\epsilon(t) + 2\big) \\
         &= 2\epsilon(t) + \epsilon^2(t)
\end{align*}
Thus, for $\epsilon(t) \ll 1$, the maximum Euclidean error for an arc of length $\phi$ radians on the unit circle is
\\[-20pt]
\begin{align*}
\epsilon_{max}(\phi) \approx \frac12 \psi_{max}(\phi)
                            = \frac{2}{27}\frac{\sin^6 \frac\phi4}{\cos^2\frac\phi4}
\end{align*}

\begin{table}[H]
\begin{mdframed}[hidealllines=true,backgroundcolor=gray!6]
\centering
\begin{tabular}{ |r l|l l| }
 \hline
 \multicolumn{2}{|c|}{\rule{0pt}{2.5ex} \textbf{Arc length $\bm{\phi}$ (radians)}} & \multicolumn{2}{|c|}{\textbf{Error $\bm{\epsilon_{max}}$ (Euclidean dist.)}} \\[.3ex] \hline \hline
 \rule{0pt}{2.5ex} $\quad\quad 0.1\pi$ & (0.314159) & $\quad\quad 0.000000017$ & (1.7E-8)$\quad$ \\[.3ex] \hline
 \rule{0pt}{2.5ex} $0.2\pi$ & (0.628319) & $\quad\quad 0.0000011$   & (1.1E-6)$\quad$  \\[.3ex] \hline
 \rule{0pt}{2.5ex} $0.3\pi$ & (0.942478) & $\quad\quad 0.000013$    & (1.3E-5)  \\[.3ex] \hline
 \rule{0pt}{2.5ex} $0.4\pi$ & (1.256637) & $\quad\quad 0.000071$    & (7.1E-5)  \\[.3ex] \hline
 \rule{0pt}{2.5ex} $0.5\pi$ & (1.570796) & $\quad\quad 0.00027$     & (2.7E-4)  \\[.3ex] \hline
 \rule{0pt}{2.5ex} $0.6\pi$ & (1.884956) & $\quad\quad 0.00082$     & (8.2E-4)  \\[.3ex] \hline
 \rule{0pt}{2.5ex} $0.7\pi$ & (2.199115) & $\quad\quad 0.0021$      & (2.1E-3)  \\[.3ex] \hline
 \rule{0pt}{2.5ex} $0.8\pi$ & (2.513274) & $\quad\quad 0.0047$      & (4.7E-3)  \\[.3ex] \hline
 \rule{0pt}{2.5ex} $0.9\pi$ & (2.827433) & $\quad\quad 0.0096$      & (9.6E-3)  \\[.3ex] \hline
\end{tabular}
\caption{Maximum error $\epsilon_{max}$ for arcs of various lengths on the unit circle}
\end{mdframed}
\end{table}

Table 1 lists the maximum Euclidean error values $\epsilon_{max}(\phi)$ for arc lengths $\phi$ on the unit circle that range from $0.1\pi$ to $0.9\pi$ radians. The maximum error varies by more than five orders of magnitude over this range, with the errors at the longer arc lengths being significantly larger than the errors at smaller arc lengths.

When a circle is affine-transformed to an ellipse, how does the error in the circular arc approximation map to the ellipse?

The left side of \mbox{Figure 8} shows a circle (dashed black line) and the piecewise B\'ezier curve approximation to the circle (solid gray line). For the purpose of illustration, the approximation errors are greatly exaggerated. The maximum errors occur at the top, bottom, and sides of the circle.

On the right side of \mbox{Figure 8}, an affine transformation has squashed the sides of the circle to form an ellipse half the width of the circle but its height remains the same. The error at the sides of the ellipse shrinks to half that at the sides of the circle. But as seen in the figure, the error at the top and bottom of the ellipse remains the same as that of the circle.

%////////////////////////////////////////////////////////////////
\begin{figure}[H]
  \centering
  \fbox{%
    \begin{tabular}{p{0.9\textwidth}}
      \centering
      \includegraphics[width=9.5cm]{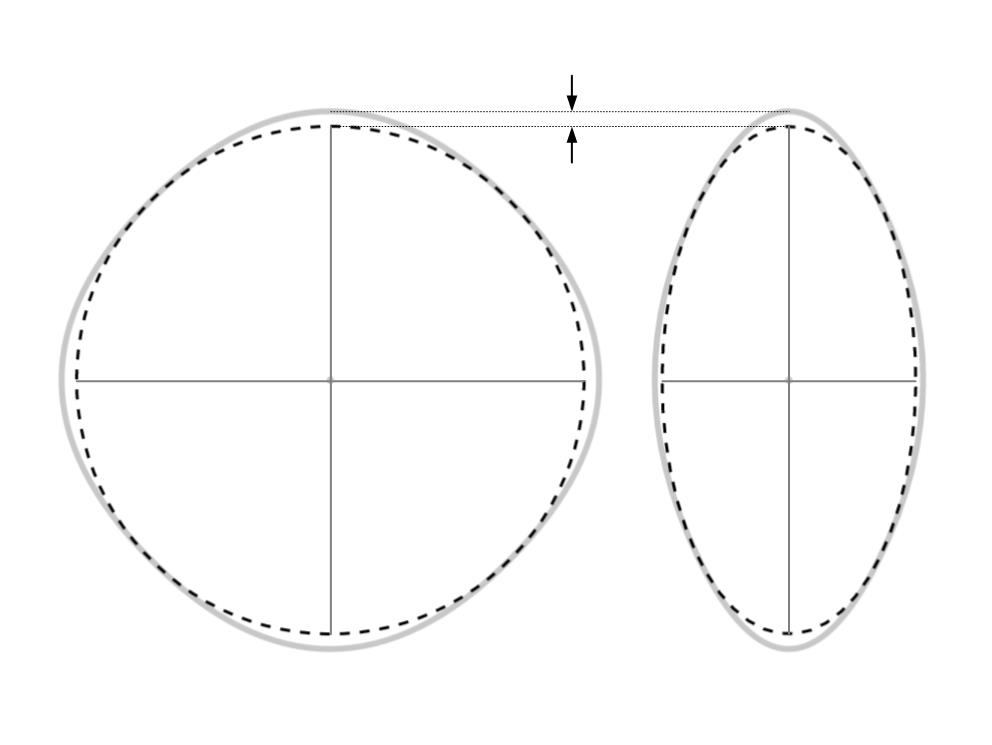} \\
      \caption{How the radial error (exaggerated here) changes when the approximate circle on the left is squashed sideways to form the ellipse on the right}
    \end{tabular}}
\end{figure}

%-------------------------------------------------------------------------------------------------------------------------------
\section{References}

\begin{enumerate}[label={[\arabic*]}]
  \item Ahn, Y. J., \& Kim, H. O. (1997). Approximation of circular arcs by Bézier curves. \textit{Journal of Computational and Applied Mathematics}, \textit{81}(1), 145-163.

  \item Apple.com. addEllipse() and addArc() functions.
In \textit{Apple developer documentation, Core Graphics}.
\textcolor{urlcolor}{\seqsplit{https://developer.apple.com/documentation/coregraphics/cgcontext/1456420-addellipse}}

  \item Bartels, R. H., Beatty, J. C., \& Barsky, B. A. (1987). \textit{An Introduction to Splines for Use in Computer Graphics \& Geometric Modeling} (pp 218-223). Morgan Kaufmann.

  \item Dokken, T., Dæhlen, M., Lyche, T., \& Mørken, K. (1990). Good approximation of circles by curvature-continuous B\'ezier curves. \textit{Computer Aided Geometric Design}, \textit7(1-4), 33-41.

  \item Farin, G. E. (2002). Some properties of B\'ezier curves. \textit{Curves and Surfaces for CAGD: A Practical Guide} (pp 47-48). Morgan Kaufmann.

  \item Foley, J. D., van Dam, A., Feiner, S. K., \& Hughes, J. F. (1990). \textit{Computer Graphics: Principles and Practice}, 2nd ed. (pp 507-509). Addison-Wesley.

  \item Goldapp, M. (1991). Approximation of circular arcs by cubic polynomials. \textit{Computer Aided Geometric Design}, \textit8(3), 227-238.

  \item Karmermans, M. (2020). Circular arcs and cubic Béziers. \textit{A Primer on Bézier Curves}. Pomax.github.io.
\textcolor{urlcolor}{\seqsplit{https://pomax.github.io/bezierinfo/\#circles\_cubic}}

  \item Leijen, D. (2015). The ellipse package. \textit{Comprehensive\,\TeX \,Archive Network}.
\textcolor{urlcolor}{https://ctan.net/graphics/ellipse/ellipse.pdf}

  \item Microsoft.com. Direct2D ArcSegment function. In \textit{Windows app development documentation, Direct2D}.
\textcolor{urlcolor}{\seqsplit{https://learn.microsoft.com/en-us/windows/win32/api/d2d1helper/nf-d2d1helper-arcsegment}}

  \item Microsoft.com. SKPath.ArcTo Method. In \textit{Windows app development documentation, SkiaSharp}.
\textcolor{urlcolor}{\seqsplit{https://learn.microsoft.com/en-us/dotnet/api/skiasharp.skpath.arcto?view=skiasharp-2.88}}

  \item Mortensen, S. (2012). Approximate a circle with cubic Bézier curves.
\textcolor{urlcolor}{https://spencermortensen.com/articles/bezier-circle/}

  \item Mozilla.org. CanvasRenderingContext2D: ellipse() method. In \textit{MDN Web Docs}.
\textcolor{urlcolor}{\seqsplit{https://developer.mozilla.org/en-US/docs/Web/API/CanvasRenderingContext2D/ellipse}}

  \item OpenCV.org. Ellipse. In \textit{OpenCV 2.4.13.7 documentation, OpenCV API Reference, Drawing functions}. \textcolor{urlcolor}{\seqsplit{https://docs.opencv.org/2.4/modules/core/doc/drawing\_functions.html\#ellipse}}

  \item Plauger, P. J. (1992). Chapter 7: \texttt{<math.h>}. \textit{The Standard C Library}. Prentice-Hall.

  \item Riškus, A. (2006). Approximation of a cubic Bézier curve by circular arcs and vice versa.
      \textit{Information Technology and Control}, \textit{35}(4).

  \item Schneider, P., \& Eberly, D. H. (2002). \textit{Geometric Tools for Computer Graphics} (pp 98-104). Elsevier.

  \item Tran-Thong. (1983). Ellipse, arc of ellipse and elliptic spline. \textit{Computers \& Graphics}, \textit{7}(2), 169-175.

  \item Van\nobreak\hspace{.10em}Aken, J., \& Simar, R. (1988). A Conic Spline Algorithm.
      \textit{TMS34010 Application Guide} (pp 255-278). Texas Instruments Incorporated.

  \item Van\nobreak\hspace{.10em}Aken, J., \& Simar, R. (1992). A Parametric Elliptical Arc Algorithm.
      In \textit{Graphics Gems III (IBM Version)} (pp 164-172). Morgan Kaufmann.

  \item Van\nobreak\hspace{.10em}Aken, J. R. (2018). A rotated ellipse from three points. ResearchGate.
\textcolor{urlcolor}{https://www.researchgate.net/profile/Jerry\_Van\_Aken}

  \item Van\nobreak\hspace{.10em}Aken, J. R. (2020). A Fast Parametric Ellipse Algorithm. \textit{arXiv preprint arXiv:2009.03434}. \textcolor{urlcolor}{https://arxiv.org/pdf/2009.03434}

  \item W3.org. Elliptical arc implementation notes. In \textit{SVG 1.1 (Second Edition)\,$-$\,16 August 2011, Appendix F: Implementation Requirements}. \textcolor{urlcolor}{\seqsplit{https://www.w3.org/TR/SVG11/implnote.html\#ArcImplementationNotes}}

  \item W3.org. The SVG elliptical arc command. In \textit{Scalable Vector Graphics (SVG) 2, W3C Candidate Recommendation 04 October 2018}. \textcolor{urlcolor}{\seqsplit{https://www.w3.org/TR/SVG/paths.html\#PathDataEllipticalArcCommands}}

  \item Watt, A., Watt, M. (1992). \textit{Advanced Animation and Rendering Techniques} (p 69). Addison-Wesley.

\end{enumerate}

\end{document}